\documentclass[12pt]{article}
\usepackage[T1]{fontenc}
\usepackage{mathptmx}
\usepackage{titling}

\usepackage{xcolor,soul,framed} 
\colorlet{shadecolor}{yellow}
\usepackage[cmex10]{amsmath}
\usepackage{array}
\usepackage{mdwmath}
\usepackage{mdwtab}
\usepackage{eqparbox}
\usepackage{url}
\usepackage{multirow}
\usepackage{algorithmic}
\usepackage[linesnumbered,ruled,vlined]{algorithm2e}
\usepackage{multicol}
\usepackage{textcomp}
\usepackage{placeins}
\usepackage{subfig}
\usepackage{lineno}

\usepackage[english]{babel}

\usepackage[a4paper,top=2cm,bottom=2cm,left=3cm,right=3cm,marginparwidth=1.75cm]{geometry}

\usepackage{amsmath}
\usepackage{graphicx}
\usepackage[colorlinks=true, allcolors=blue]{hyperref}
\usepackage[document]{ragged2e}

\date{}
\begin{document}

\begin{flushleft}
\Large \textbf{Study of keyword extraction techniques for Electric Double Layer Capacitor domain using text similarity indexes: An experimental analysis}
\end{flushleft}
\begin{flushleft}
\large M. Saef Ullah Miah\textsuperscript{1}, Junaida Sulaiman\textsuperscript{1,}\textsuperscript{2}, Talha Bin Sarwar\textsuperscript{3}, Kamal Z. Zamli\textsuperscript{1} and Rajan Jose\textsuperscript{4} \\
\small \textsuperscript{1}Faculty of Computing, College of Computing and Applied Sciences, Universiti Malaysia Pahang, 26600, Pekan, Malaysia.\\
\small \textsuperscript{2}Center for Data Science and Artificial Intelligence (Data Science Center), Universiti Malaysia Pahang, 26600, Pekan, Malaysia.\\
\small \textsuperscript{3}Department of Computer Science, Faculty of Science and Technology, American International University-Bangladesh(AIUB), Dhaka, Bangladesh.\\
\small \textsuperscript{4}Faculty of Industrial Sciences \& Technology, Universiti Malaysia Pahang, 26300, Gambang, Malaysia.
\\
\vspace{2.5mm}
\small Corresponding author: Talha Bin Sarwar; talhasarwar40@gmail.com
\end{flushleft}


\section*{Abstract}
\justifying
Keywords perform a significant role in selecting various topic-related documents quite easily. Topics or keywords assigned by humans or experts provide accurate information. However, this practice is quite expensive in terms of resources and time management. Hence, it is more satisfying to utilize automated keyword extraction techniques. Nevertheless, before beginning the automated process, it is necessary to check and confirm how similar expert-provided and algorithm-generated keywords are. This paper presents an experimental analysis of similarity scores of keywords generated by different supervised and unsupervised automated keyword extraction algorithms with expert provided keywords from the Electric Double Layer Capacitor (EDLC) domain. The paper also analyses which texts provide better keywords like positive sentences or all sentences of the document. From the unsupervised algorithms, YAKE, TopicRank, MultipartiteRank, and KPMiner are employed for keyword extraction. From the supervised algorithms,  KEA and WINGNUS are employed for keyword extraction. To assess the similarity of the extracted keywords with expert-provided keywords, Jaccard, Cosine, and Cosine with word vector similarity indexes are employed in this study. The experiment shows that the MultipartiteRank keyword extraction technique measured with cosine with word vector similarity index produces the best result with 92\% similarity with expert provided keywords. This study can help the NLP researchers working with the EDLC domain or recommender systems to select more suitable keyword extraction and similarity index calculation techniques.

\paragraph*{Keywords--}
\textbf{Keyword similarity calculation; keyword extraction comparison; keyword validation; Electric Double Layer Capacitor; EDLC}

\section{Introduction}
Keywords are significant for automated document processing. Keywords are the concise representation of the contents of a document~\cite{Rose2010}. From keywords, the context of the documents can be easily understood. When there is a need to process lots of documents or classify any document for any purpose, it is tedious to go through the whole document one by one and classify them. Instead, going through the keywords makes this process faster, even for a human. However, it is also a time-consuming process to go through the keywords for many documents by a human. This task can be automated by employing machines to look for the keywords and classify the documents. Since the process of keyword extraction is being automated, it should also be assured that extracted keywords represent the actual context of the document; else automated extraction will be a complete loss of time and resources. This assurance can be done by comparing the extracted keywords with human or expert assigned keywords. Therefore, this paper introduces an experimental study to measure the similarity score between expert-provided keywords and keyword extraction algorithms generated keywords to observe how similar the machine-generated keywords' values are to the expert provided keywords. In other words, this experiment can guide if the machine-generated keywords are feasible to utilize instead of expert-provided keywords for any specific domain.

There are several different keyword extraction algorithms available at present \cite{hasan2014automatic,trendtw}. These algorithms are employed in different scenarios, such as recommender systems, trend analysis, similar document identification, relevant document selection~\cite{sarwar,geofencing,miah}. All these algorithms are divided into three primary categories based on their extraction technique: supervised, unsupervised, and semi-supervised technique~\cite{Beliga}. This study compares the similarity scores for supervised and unsupervised techniques with three prominent similarity indexes, namely, Jaccard similarity index~\cite{Jaccard1912}, Cosine similarity index~\cite{cosine1,cosine2} and Cosine with Word vector similarity~\cite{mikolov2013efficient}. The key contributions of this work are,
\begin{itemize}
	\item Recommending a keyword extraction technique that provides more similar machine-generated keywords to the expert or human provided keywords.
	\item Recommending type of texts (positive texts only or whole text of a document) that provides more similar keywords.
	\item Recommending a better similarity index for measuring similarity score between documents.
	\item Finding the feasibility of utilizing machine-generated keywords instead of expert-curated keywords.
\end{itemize}

The rest of the paper is organized as follows. Employed keyword extraction techniques and relevant works are presented in Section~\ref{sec:bs} with their known shortcomings and strengths. Employed methodologies for the experiment are mentioned in Section~\ref{sec:method}. Then, the result analysis of the experiment is discussed in Section~\ref{sec:exp_eva}, and concluding remarks in Section~\ref{sec:con}.

\section{Background Study}
\label{sec:bs}
In this paper, some notable and well-known similarity index calculation algorithms and keyword extraction algorithms are employed. All the text-similarity and keyword extraction algorithms with shortcomings and strengths are discussed in this section.

\subsection{Keyword Extraction}
Keyword extraction from text is an analysis technique that automatically extracts the most used and most important words or phrases from text based on different parameters~\cite{Firoozeh2020}. In some techniques, these parameters can be defined externally, and some techniques do not support external definition~\cite{Beliga}.  Mainly there are three classes of keyword extraction techniques. Among them, supervised and unsupervised techniques are employed in this study.

\subsubsection{Unsupervised Keyword Extraction}
Four unsupervised keyword extraction techniques are employed in this paper. Unsupervised techniques are prone to poor accuracy and require a larger corpus input, and do not extrapolate well~\cite{bennani-smires-etal-2018-simple}. However, unsupervised techniques are utilized widely compared to supervised techniques, as all sorts of domain-specific training labeled data are not always available for all the domains.  

\paragraph{YAKE}
YAKE was proposed by Campos et al.,~\cite{Campos2020}. It is a lightweight unsupervised keyword extraction technique based on TF-IDF. YAKE extracts keywords by calculating five features, namely, Word Casing (WC), Word position (WP), Word Frequency (WF), Word Relatedness to Context (WRC), and Word DifSentence (WF). The relation between five features can be expressed through the following Equation \ref{eq:yake}, where S(w) is the measure for each word. After calculating the measure for each word, the final keyword is calculated utilizing a 3-gram model~\cite{Sun2020}.
\begin{equation}
	\label{eq:yake}
	\mathrm{S(w)=\ }\frac{WR*WP}{WC+\frac{WF}{WRC}+\frac{WD}{WR}}
\end{equation}

\paragraph{TopicRank}
Bougouin et al. proposed topicRank~\cite{bougouin2013topicrank} in 2013, which is a clustering-based model. It divides the document into multiple topics employing the hierarchical agglomerative clustering~\cite{medelyan2009human}. Then utilizing the PageRank~\cite{pagerank}, it scores each topic and selects each top-ranked candidate keyword from each topic. After that, it selects all the top candidate words as final keywords.

\paragraph{MultipartiteRank}
MultipartiteRank is a topic-based keyword extraction model. It encodes topical information of a document in a multipartite graph structure. This technique represents candidate keywords and topics of a document in a single graph, and utilizing the mutually reinforcing relationship of the candidate keywords and topics improves candidate ranking. This method has two steps of selecting candidate words as keywords, $i)$ Representing the whole document in a graph and $ii)$ Assigning relevance score to each word. Between these two steps, position information is captured utilizing edge weights adjustment. As a result, most of the time, it outperforms different other key-phrase extraction techniques~\cite{boudin-2018-unsupervised}.

\paragraph{KPMiner}
El-Beltagy et al. proposed the KP-miner~\cite{el2009kp} in 2009. This method also utilizes TF-IDF to calculate words as keywords. This calculation is done in three steps, $i)$ Selecting candidate words from the document utilizing least allowable seen frequency (lasf) factor and CutOff factor, $ii)$ Calculating candidate word's score, and $iii)$ Selecting the candidate word with the highest score utilizing the candidate word position and TF-IDF score as the final keyword.

\subsubsection{Supervised Keyword Extraction}
While unsupervised algorithms do not need a large amount of labeled training data, supervised algorithms need a large amount of that data and perform poorly except in the training domain. However, for any specific domain, supervised techniques are preferred over unsupervised techniques~\cite{Sun2020}. In this paper, two supervised techniques are employed, KEA and WINGNUS.

\paragraph{KEA}
KEA is a supervised keyword extraction algorithm proposed by Witten et al. in 1999~\cite{witten1999c}. KEA classifies a candidate keyword utilizing word frequency and position of the word in the document. After that, it predicts which candidate words are qualified as keywords utilizing the Naive Bayes machine learning algorithm. The machine learning model builds a predictive model initially. Then, keywords are extracted utilizing this predictive model~\cite{witten2005kea}.

\paragraph{WINGNUS}
This supervised keyword extraction technique is developed focusing on keyword extraction from scientific documents~\cite{Nguyen2010}. It utilizes inferred document logical structure~\cite{mao2003document}  in the candidate word identification process to limit the phrase number in the candidate word list. This method utilizes regular expression rules to extract candidate words, and instead of whole document text, it utilizes input text in different levels like title and headers or abstract and introduction. Like KEA, it also utilizes the Naive Bayes machine learning algorithm to select candidate words.

\subsection{Text Similarity Index}
Determining how similar two pieces of text are to each other is the simple idea of text similarity index or text similarity calculation. In this study, keywords from different documents extracted by keyword extraction algorithms and expert-provided keywords' similarity are measured. In two ways, this similarity can be measured, one is lexical similarity, and another is semantic similarity~\cite{Maheshwari2017,Yang,Aryal2019,Sitikhu2019,Thada,Steinberger2002}. This paper implemented both the similarity measures utilizing Jaccard, Cosine, and Cosine with word vector similarity indexes and presented the outcome for EDLC based scientific articles.

\subsubsection{Jaccard Similarity}
Jaccard similarity index is a lexical similarity index method, which calculates the similarity index at the word level. As lexical similarity is unaware of the word's actual meaning or the entire phrase, Jaccard similarity takes two sets of text and calculates the similarity between all pairs of sets. Jaccard provides a similarity score with a range of 0\% to 100\%. This algorithm is very sensitive to sample size and may provide unexpected results for a small sample size. Conversely, for larger sample sizes, it is computationally costly~\cite{jaccard1,jaccard2}. Jaccard similarity index is calculated utilizing the Equation \ref{eq:1}, where A and B are two different sets of text or documents.
\begin{equation}
	\label{eq:1}
	J(A,B)= \frac{|A\bigcap B|}{|A|+|B|-|A\bigcup B|}
\end{equation}

\subsubsection{Cosine Similarity}
The cosine similarity index measures the similarity between two documents utilizing the cosine angle between two multi-dimensional vectors in a multi-dimensional space regardless of their size. In this technique, sentences are converted into vectors utilizing the bag of words method. Then employing the Equation \ref{eq:2}, where A and B are two documents converted into vectors. This algorithm is computationally expensive for larger data sample\cite{cosine1,cosine2}. 

\begin{equation}
	\label{eq:2}
	\cos ({A},{ B}) = \frac{ \sum_{i=1}^{n}{{ A}_i{ B}_i} }{ \sqrt{\sum_{i=1}^{n}{({ A}_i)^2}} \sqrt{\sum_{i=1}^{n}{({ B}_i)^2}} }
\end{equation}

\subsubsection{Word Vector}
Word vectors are a type of word embedding, where similar meaningful words are arranged in a similar representation, mostly with vectors. Each word is mapped to a vector in a predefined vector space~\cite{wv}. It is different from Jaccard similarity in the way that Jaccard measures lexical similarity, but in word vector, it is measured for semantic similarity. Utilizing word vectors, similar meaningful words can be measured rather than the exact word, enabling better scores for similarity measures. In this study, as a word vector model, Wod2vec~\cite{mikolov2013efficient} proposed by Mikolov et el., is utilized. Word2vec is different from the traditional tf-idf measure, where tf-idf sets one number per word, but Word2vec sets one vector per word.

\section{Methodology}
\label{sec:method}
\begin{figure}[!htbp]
	\centering
	\includegraphics[scale=.5] {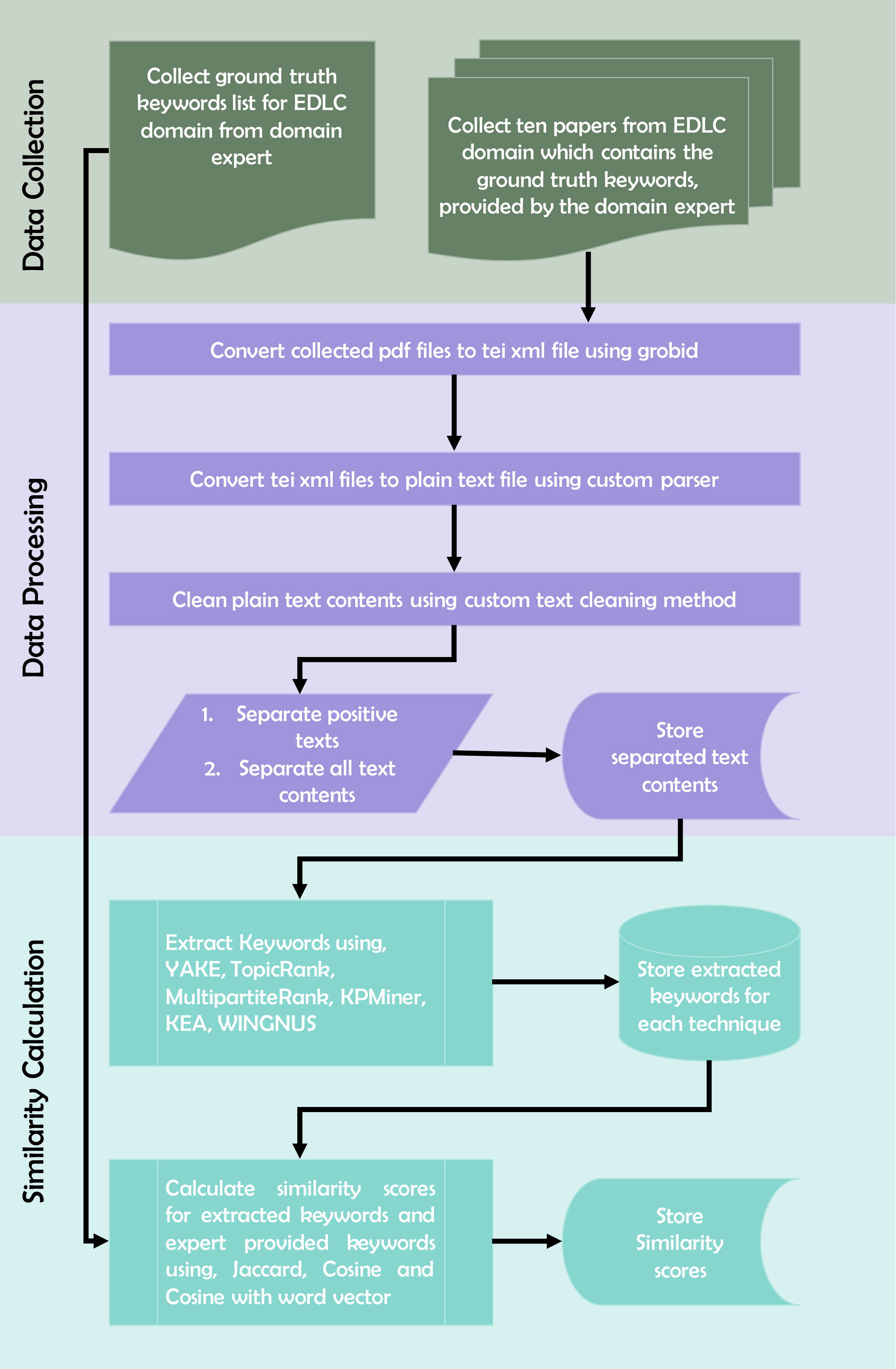}
	\caption{Overview of the employed methodology }
	\label{fig:method}
\end{figure}

This study diverges into three major components, $i)$ Data collection, $ii)$ Data Processing, and $iii)$ Similarity score calculation. In the data collection component, ground truth data and test data are collected from respective sources. Collected data is cleaned and processed for the similarity calculation component is done in the data processing component. In the similarity score calculation component, similarity scores for collected data are calculated with different similarity indexes employing different keyword extraction techniques. The conceptual overview of the employed methodology can be found in Figure \ref{fig:method}.
\subsection{Data Collection}
In this study, the Electric Double Layer Capacitor (EDLC) domain is considered as the experiment's use case. Hence, from the domain experts, a set of 32 keywords of the EDLC domain has been collected as ground truth keywords, and ten scientific documents are collected from the same domain, which satisfies the keywords and is suggested as the relevant document to the domain. The experiment is based on the quest that, from these ten documents, keywords are extracted through different keyword extraction techniques, and then extracted keywords are compared for the similarity score with the domain expert provided keywords. First column from the left of Table~\ref{tab:kw-table} contains the domain expert provided keywords for the EDLC domain. All the scientific documents are collected in portable document format (pdf), and keywords are collected in plain text.

\begin{table}[!hp]
\caption{Domain expert curated keywords for EDLC domain with lemmatised and stemmed version. From left, Keywords column contains the original keywords provided by the domain experts. Lemmatised keyword and Stemmed keyword columns contain lemmatised and stemmed version of the original keywords.}
\label{tab:kw-table}
\resizebox{\textwidth}{!}{\begin{tabular}{lll}
\hline
\textbf{Keyword}               & \textbf{Lemmatised keyword}             & \textbf{Stemmed keyword}                \\ \hline
supercapacitors                 & supercapacitors                 & supercapacitors                 \\
scs                             & sc                              & sc                              \\
electrochemical capacitors      & electrochemical capacitors      & electrochemical capacitor       \\
energy storage device           & energy storage device           & energy storage devic            \\
electric double layer capacitor & electric double layer capacitor & electric double layer capacitor \\
edlc                            & edlc                            & edlc                            \\
pseudocapacitance               & pseudocapacitance               & pseudocapacit                   \\
electrostatic adsorption        & electrostatic adsorption        & electrostatic adsorpt           \\
electrosorption                 & electrosorption                 & electrosorption                 \\
faradaic redox reactions        & faradaic redox reactions        & faradaic redox react            \\
stern layer                     & stern layer                     & stern lay                       \\
helmholtz double layer          & helmholtz double layer          & helmholtz double lay            \\
double layer formation          & double layer formation          & double layer formation          \\
activated carbon                & activated carbon                & activated carbon                \\
porous carbon                   & porous carbon                   & porous carbon                   \\
carbon nanotubes                & carbon nanotubes                & carbon nanotubes                \\
graphene                        & graphene                        & graphene                        \\
graphite oxide                  & graphite oxide                  & graphite oxide                  \\
go                              & go                              & go                              \\
reduced graphite oxide          & reduced graphite oxide          & reduced graphite oxide          \\
rgo                             & rgo                             & rgo                             \\
surface charge accumulation     & surface charge accumulation     & surface charge accumul          \\
high power applications         & high power applications         & high power appl                 \\
charge separation at electrode interface     & charge separation at electrode interface   & charge separation at electrode interfac   \\
charge separation at electrolyte   interface & charge separation at electrolyte interface & charge separation at electrolyte interfac \\
non-faradaic process            & non-faradaic process            & non-faradaic process            \\
specific surface area           & specific surface area           & specific surface area           \\
pore size distribution          & pore size distribution          & pore size distribut             \\
electrochemical interface       & electrochemical interface       & electrochemical interfac        \\
edlc characteristics            & edlc characteristics            & edlc characterist               \\
diffuse double layer            & diffuse double layer            & diffuse double lay              \\
polarizable capacitor electrode & polarizable capacitor electrode & polarizable capacitor electrod  \\ \hline
\end{tabular}}
\end{table}



\subsection{Data Processing}
\label{subsec:dp}
In the data processing stage, collected pdf files are initially converted to plain text format. To convert the files, grobid~\cite{GROBID} tool is utilized, which primarily converts the pdf files to tei xml format and then with a custom tei xml parser xml contents are converted to a plain text file. The custom xml parser is developed by the authors utilizing the python programming language. After the conversion, text contents are cleaned to remove extra spaces, special characters, extra line breaks, parentheses, references, figures, and tables employing a custom data cleaning method also developed by the authors.

Text cleaning methods are dependent on the dataset and desired output. However, apart from the dataset and output, several steps are commonly performed to clean text data, namely removing punctuation, filtering out stop words, stemming and lemmatisation, and converting text to upper and lower case. For the dataset used in this study, some of the common cleaning tasks are implemented, and some of them are avoided. In addition to these tasks, some dataset-specific cleanup tasks are also performed. Based on the cleanup activities performed in the dataset, the cleaning process is described as a custom text cleaning process. For example, normalization of non-standard words (NSW) is not performed in the text cleaning process. NSW are words that are not available in a dictionary, such as numbers, dates, abbreviations, chemical symbols of materials, currency amounts, and acronyms~\cite{nsw}. Most scientific papers contain these NSWs, and they refer to specific processes or operations of any domain which are not available on a dictionary, e.g., “MnO2”, a chemical symbol for a material called Manganese dioxide. Stemming and lemmatisation operations on the words are also discarded since most keywords are a combination of several words, e.g., “helmholtz double layer”, which gives the same result when lemmatised and a meaningless result when stemmed.  Table \ref{tab:kw-table} represents the original keywords with the lemmatised and stemmed version of the keywords. From Table \ref{tab:kw-table}, it can be observed that the output of the lemmatised keywords is almost similar to the original keywords, and the stemmed version of the keywords produces unintelligible words. In the dataset-specific cleaning process, all tabular data, references, and images are removed from the articles. Then, the text contents are decoded from the UTF8 encoding format. In addition to normalizing these decoded text contents, some special character substitution operations are performed.

\begin{figure}[!htbp]
	\centering
	\includegraphics[scale=.70] {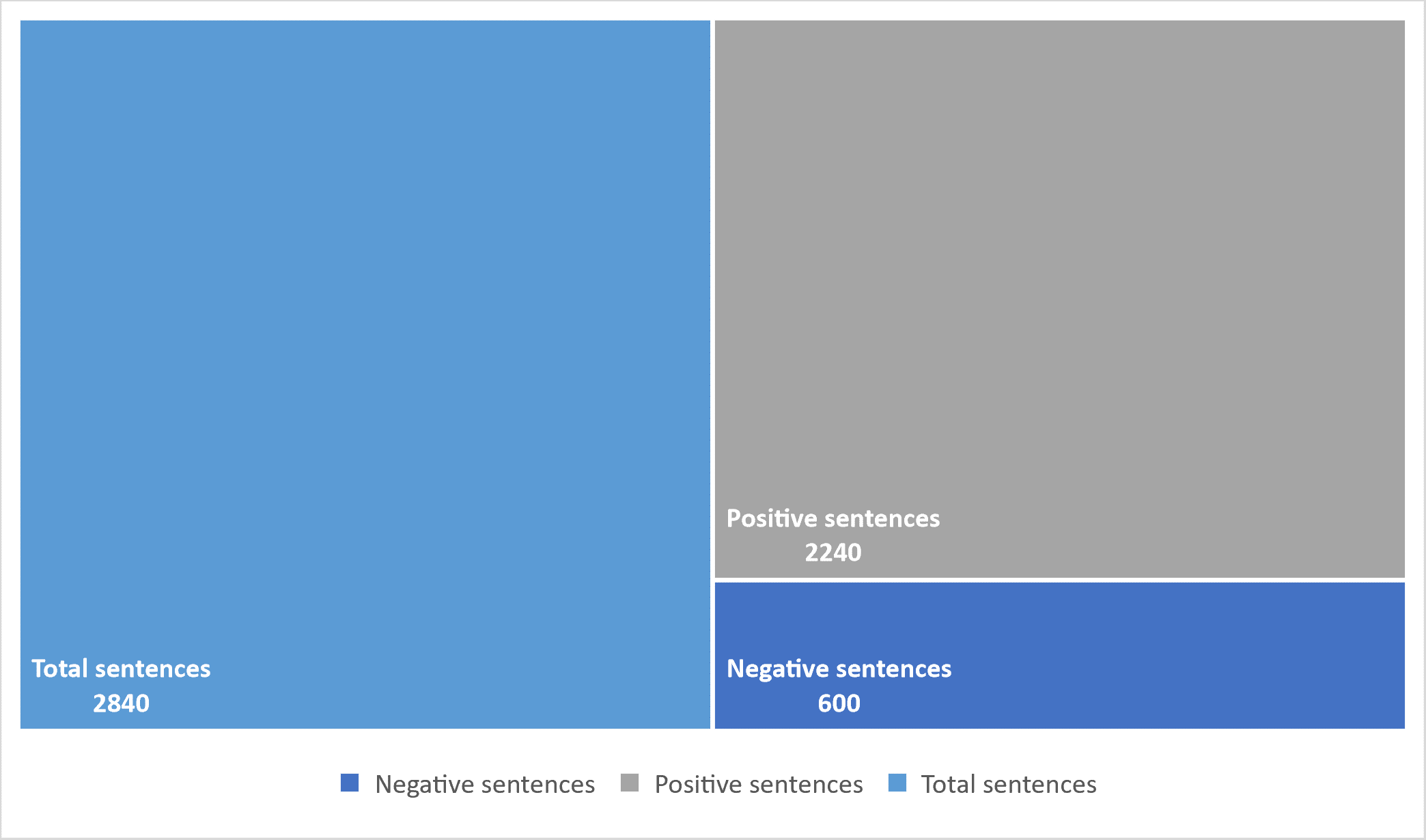}
	\caption{Positive and Negative sentence distribution of the dataset utilized in this study.}
	\label{fig:dataset}
\end{figure}

Then, from the cleaned text of each document, texts are separated into positive sentences only and all text of the document. For each document, these two types of texts are stored for the similarity calculation component. Positive sentences are identified utilizing negatives and negation–grammar Rules~\cite{Grammarly2021,Col1998,HaCohen-Kerner2016}. There are 2840 sentences in the dataset utilized in this study. Among 2840 sentences, 2240 sentences are positive sentences. Figure \ref{fig:dataset} represents the overview of the dataset stating the number of total, positive and negative sentences. The dataset can be requested through the github repository, \url{https://github.com/ping543f/kwd-extraction-study}

\subsection{Similarity Calculation}
\label{ssec:simcal}
With two sets of text obtained from the data processing component, all keyword extraction algorithms are employed to extract keywords from each set of each document. Firstly texts are passed into all the keyword extraction techniques, namely YAKE, TopicRank, MultipartiteRank, KPMiner, KEA, and WINGNUS. All techniques return the extracted keywords of the provided texts of a document. Then those keywords and expert-provided keywords are passed to the similarity index calculator to calculate the similarity score between them. Three similarity indexes are utilized to calculate the similarity score, namely, Jaccard, Cosine, and Cosine with word vector similarity index. This whole process is executed for all the documents with positive and all texts of each document. After processing each document, scores are stored with appropriate labels to analyze the result. The similarity calculation component for the scenario described above can be expressed through the Algorithm \ref{algo_1} provided below.

\begin{algorithm}[!htbp]
	
	\LinesNumbered
	
	\SetAlgoLined
	
	\caption{Similarity Score Calculation}
	\label{algo_1}
	\SetKwInOut{Input}{Input}
	\SetKwInOut{Output}{Output}
	\SetKwFunction{Fgs}{get\_score}
	\SetKwFunction{Main}{main}
	\Input{Whole text String $A\_string$}
	\Input{Positive sentence String $P\_string$}
	\Input{Domain expert curated keywords list's string $KW\_string$}
	\Output{ String containing filename, algorithm and score}
	
	\SetKwProg{Fn}{Def}{:}{}
	\Fn{\Fgs{$Sim\_algo$,$KPalgo\_name$, $text\_content$, $KW\_string$}}{
		$score$ = 0\\
		$algo\_list$ = ["yake","topicrank","multipartiterank",\\"kpminer","kea","wingnus"]\\
		\If{$KPalgo\_name$ in $algo\_list$}{
			$algo\_name$ = $KPalgo\_name$
			$keyWords$ = Extract Keywords using $algo\_name$ algorithm from $text\_content$ \\
			$SimScore$ = Calculate similarity of $keyWords$ with $KW\_string$ using $Sim\_algo$\\
			$score$ = $SimScore$\\
			\KwRet $score$\\   
		}
		\Else{
			\KwRet $error\_msg$\\
	}}
	\Fn{\Main{$Kw\_args$}}{       
		$sim\_algo$ = [jaccard, cosine, coswv]\\
		$algorithm\_list$ = ["yake", "topicrank", "multipartiterank", "kpminer", "kea", "wingnus"]\\
		\For{$algo$ in $sim\_algo$}{
			\For{$algorithm$ in $algorithm\_list$}
			{
				$score\_a$ = get\_score($algo$,$algorithm$, $A\_string$, $KW\_string$)\\
				$score\_p$ = get\_score($algo$,$algorithm$, $P\_string$, $KW\_string$)\\
				$r\_string$ = $algo$ + $algorithm$ + $score\_a$ + $score\_p$
				
			}
			\KwRet $r\_string$
		}
	}

\end{algorithm}

\subsection {Experimental Setup}
All experiment-related codes are developed utilizing Python programming language version 3.7.3~\cite{10.5555/1593511} for this study. Jaccard and Cosine similarity algorithms are developed following the equation described in their original papers~\cite{Jaccard1912,Salton1988}. Cosine similarity with word vector algorithm is implemented utilizing Spacy Python library~\cite{spacy}. All keyword extraction algorithms are implemented utilizing pke~\cite{boudin:2016:COLINGDEMO} Python package. The experiment is done in a MacBook with macOS Big Sur operating system version 11.5 with a 1.2 GHz dual-core Intel Core m5 processor and 8 gigabytes of RAM.

\section{Results and Discussion}
\label{sec:exp_eva}
\begin{figure}[!thbp]
    \centering
    \subfloat[Similarity score distribution of Positive and All sentences for unsupervised YAKE, TopicRank, MultipartiteRank and KPMiner keyword extraction algorithms for all the similarity indexes.]{
        \hspace*{-.1in}
        \includegraphics[scale=0.25]{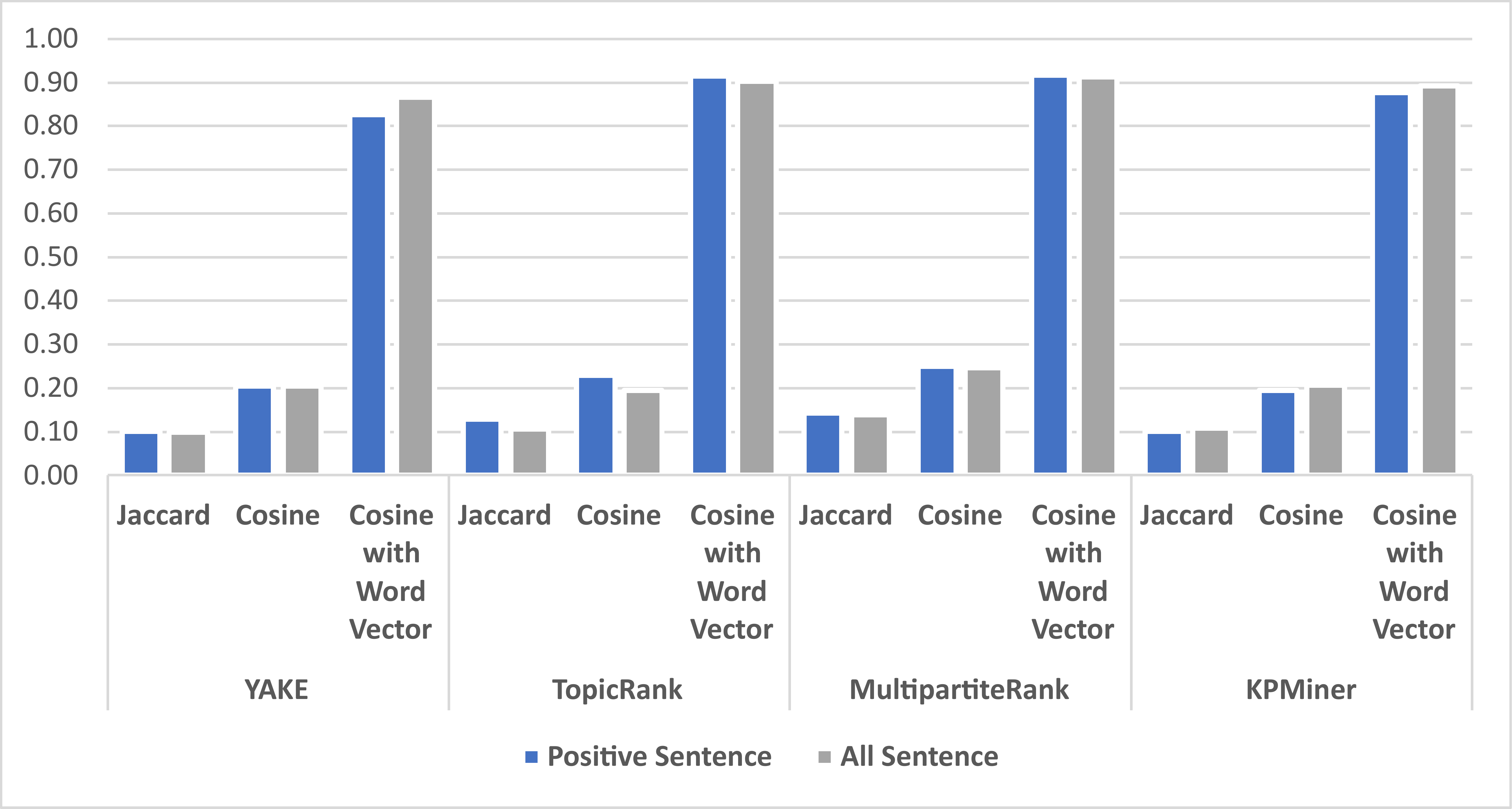}
        \label{fig:unsup}
    }
    \hfill
    \subfloat[Similarity score distribution of Positive and All sentences for supervised KEA and Wingnus keyword extraction algorithms for all the similarity indexes.]{
        \hspace*{-.1in}
        \includegraphics[scale=0.25]{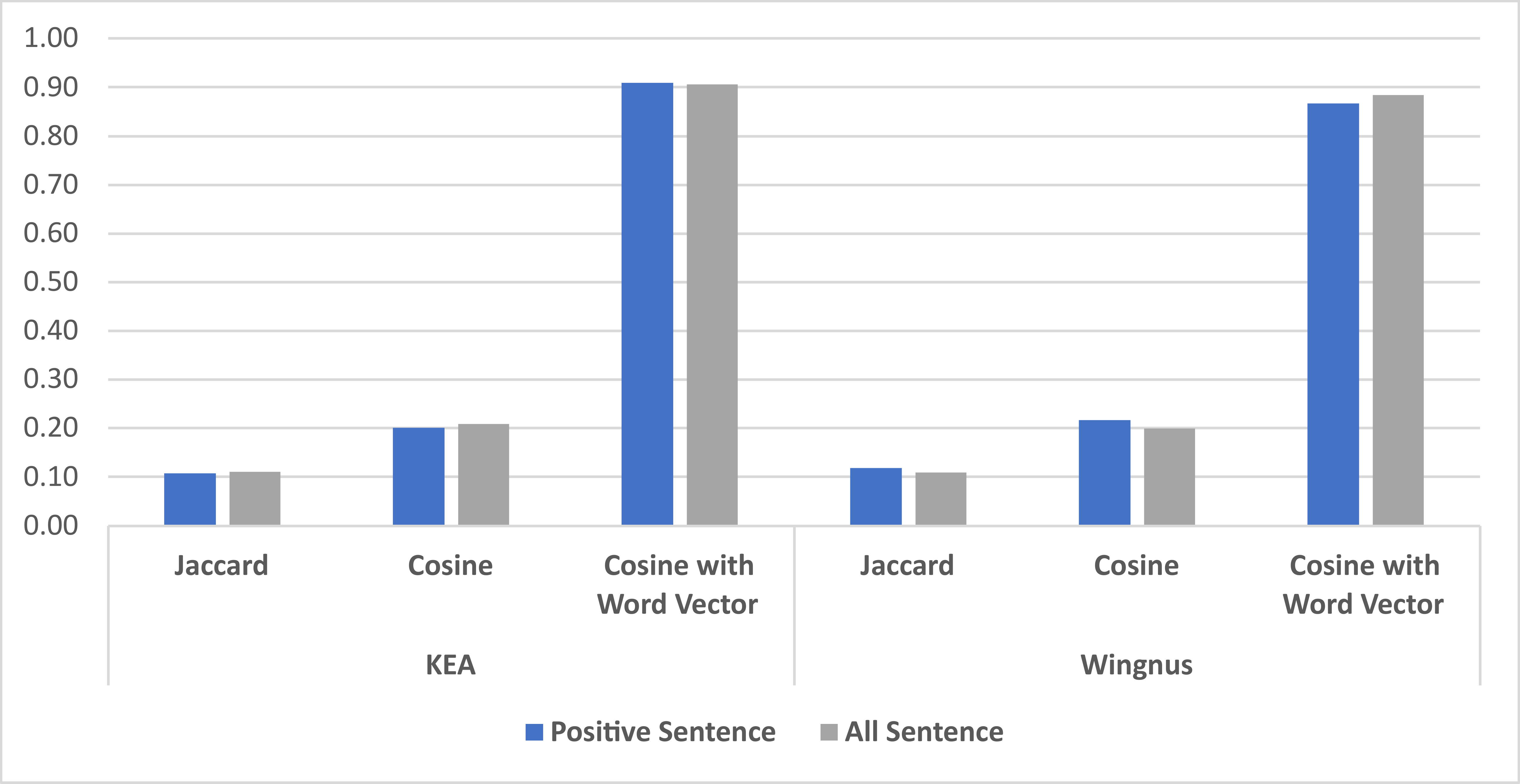}
        \label{fig:sup}
    }
    \caption{Distribution of similarity scores of supervised and unsupervised keyword extraction techniques employed in positive and all sentences for Jaccard, Cosine and Cosine with Word Vector similarity indexes.}
    \label{fig:distribution}
\end{figure}

To begin with the result analysis, Table \ref{tab:unsupervised} and Table \ref{tab:supervised} are generated from the experiment. Both tables contain the similarity scores of ten standard documents generated by different keyword extraction techniques and similarity index algorithms. Table \ref{tab:unsupervised} contains the results obtained from the unsupervised keyword extraction techniques, and Table \ref{tab:supervised} contains the results generated by the supervised keyword extraction techniques. For unsupervised techniques, the MultipartiteRank algorithm performs better in all three similarity indexes than other implemented keyword extraction techniques. Furthermore, it gives the best result of 92\% similarity score for positive sentences and 91\% for all sentences of the documents while employed with the Cosine with word vector similarity index. The lowest performing similarity index algorithm is the Jaccard similarity index for the same keyword extraction technique with a score of 14\% similarity score for both positive and all sentences of the documents. It is also observed from the experimental result that, Cosine with word vector similarity index is consistently performing better than Jaccard and cosine similarity index for all the unsupervised keyword extraction techniques. This analysis can easily be understood from Figure \ref{fig:unsup}. This figure presents the distribution of all the similarity scores of all the unsupervised techniques employed in this study for Jaccard, Cosine, and Cosine with word vector similarity indexes.

\begin{table}[!htbp]
	\centering
	\caption{Similarity scores calculated for different unsupervised keyword extraction techniques.}
	\label{tab:unsupervised}
	\begin{tabular}{cccc}
		\hline
		\multicolumn{4}{c}{\textbf{YAKE}} \\ \hline
		& \textit{\textbf{Jaccard}} & \textit{\textbf{Cosine}} & \textit{\textbf{Cosine with Word   Vector}} \\ \hline
		\textbf{Positive   Sentence} & 0.10 & 0.20 & 0.83 \\ 
		\textbf{All Sentence} & 0.10 & 0.21 & 0.87 \\ \hline
		\multicolumn{4}{c}{\textbf{TopicRank}} \\ \hline
		& \textit{\textbf{Jaccard}} & \textit{\textbf{Cosine}} & \textit{\textbf{Cosine with Word   Vector}} \\ \hline
		\textbf{Positive Sentence} & 0.13 & 0.23 & 0.91 \\ 
		\textbf{All Sentence} & 0.11 & 0.19 & 0.90 \\ \hline
		\multicolumn{4}{c}{\textbf{MultipartiteRank}} \\ \hline
		& \textit{\textbf{Jaccard}} & \textit{\textbf{Cosine}} & \textit{\textbf{Cosine with Word   Vector}} \\ \hline
		\textbf{Positive Sentence} & 0.14 & 0.25 & 0.92 \\ 
		\textbf{All Sentence} & 0.14 & 0.25 & 0.91 \\ \hline
		\multicolumn{4}{c}{\textbf{KPMiner}} \\ \hline
		& \textit{\textbf{Jaccard}} & \textit{\textbf{Cosine}} & \textit{\textbf{Cosine with Word   Vector}} \\ \hline
		\textbf{Positive Sentence} & 0.10 & 0.19 & 0.88 \\ 
		\textbf{All Sentence} & 0.11 & 0.21 & 0.89 \\ \hline
	\end{tabular}
\end{table}

\begin{table}[!htbp]
	\centering
	\caption{Similarity scores calculated for different supervised keyword extraction techniques.}
	\label{tab:supervised}
	\begin{tabular}{cccc}
		\hline
		\multicolumn{4}{c}{\textbf{KEA}} \\ \hline
		& \textit{\textbf{Jaccard}} & \textit{\textbf{Cosine}} & \textit{\textbf{Cosine with Word   Vector}} \\ \hline
		Positive   Sentence & 0.11 & 0.20 & 0.91 \\ 
		All Sentence & 0.11 & 0.21 & 0.91 \\ \hline
		\multicolumn{4}{c}{\textbf{Wingnus}} \\ \hline
		& \textit{\textbf{Jaccard}} & \textit{\textbf{Cosine}} & \textit{\textbf{Cosine with Word   Vector}} \\ \hline
		Positive   Sentence & 0.12 & 0.22 & 0.87 \\ 
		All Sentence & 0.11 & 0.20 & 0.88 \\ \hline
	\end{tabular}
\end{table}

On the other hand, for the supervised techniques, the KEA keyword extraction algorithm performs the best with 91\% of similarity score while calculating with the Cosine with word vector similarity index for both positive and all sentences of the documents. However, the WINGNUS supervised keyword extraction technique provides better similarity scores for Cosine and Jaccard similarity indexes only for positive sentences, which are 22\% and 12\% similarity scores. Nevertheless, KEA is performing better for all sentences while measured with Jaccard and Cosine similarity indexes. However, KEA holds the best similarity score utilizing the Cosine with word vector similarity index, which is around 70\% more than those measured with Jaccard and Cosine similarity index. This analysis can be more clear with a visual representation. Figure \ref{fig:sup} represents the distribution of all the similarity scores for all the supervised keyword extraction techniques with all three similarity indexes. 

\begin{figure}[!htbp]
	\centering
	\includegraphics[width=\linewidth] {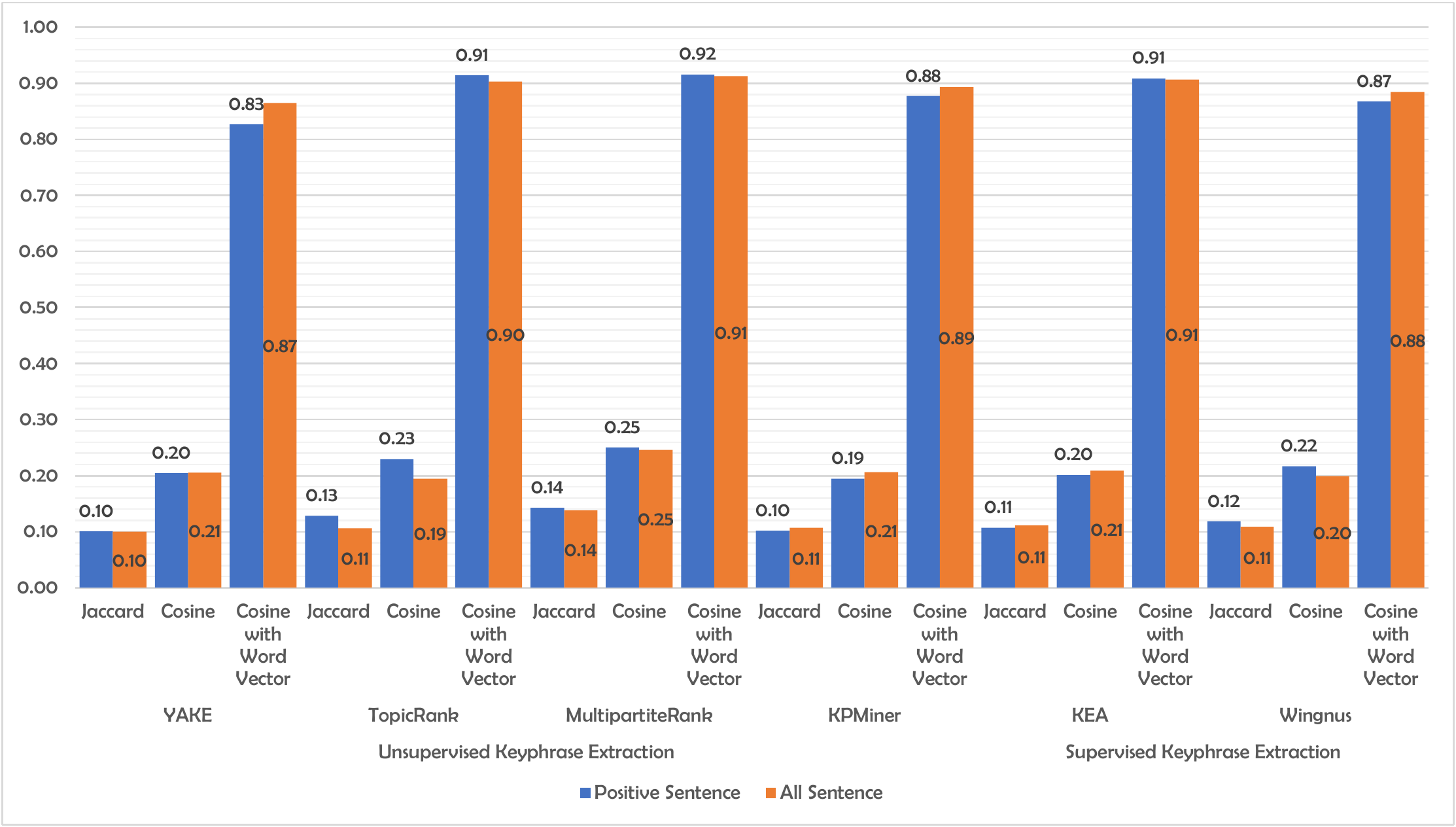}
	\caption{Similarity scores of different supervised and unsupervised keyword extraction techniques for Jaccard, Cosine and Cosine with Word vector similarity indexes.}
	\label{fig:sup-unsup}
\end{figure}

Among supervised and unsupervised keyword extraction techniques, the unsupervised technique, namely, MultipartiteRank, exhibits better performance in achieving a higher similarity score for positive sentences while measured with Cosine with word vector similarity index. Furthermore, for all sentences, unsupervised technique, MultipartiteRank, and supervised technique, KEA produces the same score of 91\% in Cosine with word vector similarity index. Similarity score comparisons for both supervised and unsupervised methods are projected in Figure \ref{fig:sup-unsup}. 

Since there are two sets of textual data: Data with positive sentences and Data with all sentences, they have implications for the experimental results seen in Table \ref{tab:unsupervised} and Table \ref{tab:supervised}. The initial hypothesis of having two separate text datasets from the same articles is to observe how positive and negative sentences affect the similarity score of the extracted keywords with the keywords provided by the experts for the specific domain, and based on this impact, recommend the relevant text data to be used. From the experimental results, the positive sentences have a minimal impact on the similarity scores for all three similarity indices compared to the scores for all sentences. This is because the negative sentences contain very few to no keywords that could match the keywords given by the experts. Therefore, there is no or minimal effect of the similarity indices between the positive sentences and the dataset with all sentences, as shown in the experimental result. The similarity values between the positive sentences and all sentences vary from 1\% to 4\%. For example, in the MultipartiteRank algorithm, the Jaccard and Cosine similarity values are the same for both texts, 14\% and 25\%, respectively. However, for the cosine with word vector similarity index, the text of the positive sentence achieves 92\% similarity, and the text of all sentences achieves 91\% similarity, which is a minimal difference of 1\%. On the other hand, in the algorithm KEA, the similarity value of cosine with word vector is the same for both text data, i.e., 91\% of similarity value. The maximum difference of 4\% in similarity score is observed for the YAKE algorithm in similarity index Cosine with Word Vector. Hence, it can be said that positive sentences and all sentences have a similar effect on the similarity index with very little difference from 1\% to 4\%.

\begin{figure}[!htbp]
	\centering
	\includegraphics[scale=.25] {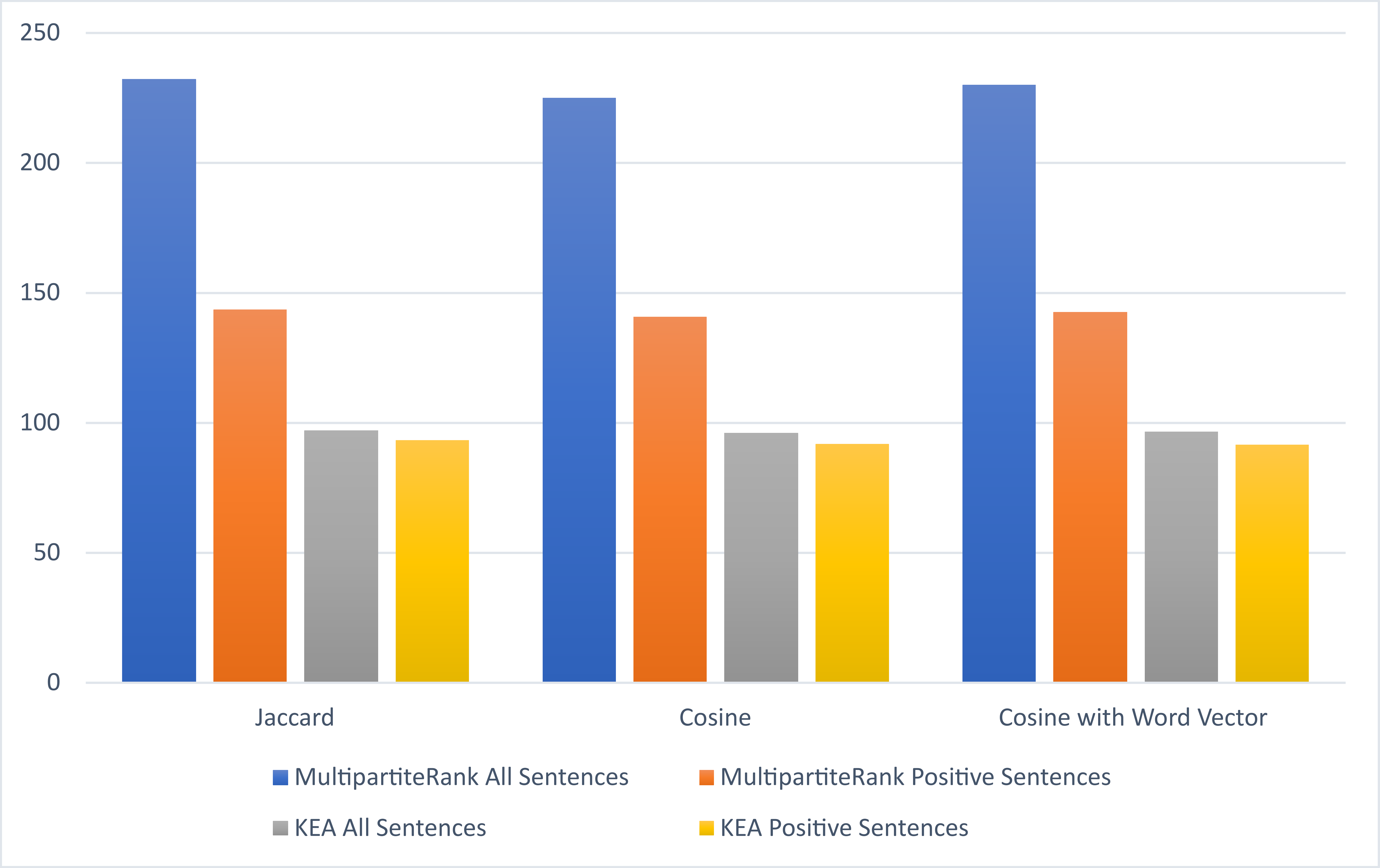}
	\caption{Comparative scores of similarity calculation run-times for positive and all sentences employing MultipartiteRank and KEA keyword extraction algorithms.}
	\label{fig:runtime}
\end{figure}

Although the positive sentences have a negligible effect on the similarity computation, they have a more significant impact on the running time of the similarity computation process. From the experiment results, the unsupervised algorithms MultipartiteRank and the supervised algorithms KEA perform better than the other algorithms used in terms of similarity index. Therefore, a runtime comparison is performed for both algorithms to study the runtime for both positive and all text sets for computing all similarity indices. Table \ref{tab:runtime} presents the runtime comparison result for the two better-performing keyword extraction techniques MultipartiteRank and KEA for Jaccard, Cosine, and Cosine similarity with word vector indices. The runtimes reported in the table \ref{tab:runtime} are the average of 5 runtimes of the experiment, which includes only the similarity computation. From the runtime table, it can be seen that positive texts have a great impact on the duration of the similarity calculation. When computing the similarity of the texts with the keywords given by the experts, the positive sentences take significantly less time than computing the similarity of all sentences. For example, in the unsupervised MultipartiteRank algorithm, the computation of all sentences takes 232.4, 225.1, and 230.2 seconds for the Jaccard, Cosine, and Cosine with word vector similarity indices, respectively. On the other hand, the computation of positive sentences takes only 143.6, 140.86, and 142.7 seconds for Jaccard, Cosine, and Cosine with word vector similarity indices, respectively, which is 88.8, 84.24, and 87.5 seconds less for the aforementioned similarity indices. A similar pattern is also observed for the supervised KEA algorithm, i.e., computing the similarity of positive sentences takes less time than computing all sentences. Figure \ref{fig:runtime} shows the comparison results in a more understandable form.

\begin{table}[!htbp]
\centering
\caption{Run-time comparison in seconds(s) of positive and all sentences texts for MultipartiteRank and KEA keyword extraction algorithms in terms of Jaccard, Cosine and Cosine with Word Vector similarity indexes.}
\label{tab:runtime}
\begin{tabular}{lccc}
\hline
 & \multicolumn{1}{l}{\textbf{Jaccard}} & \multicolumn{1}{l}{\textbf{Cosine}} & 
  \multicolumn{1}{l}{\textbf{\begin{tabular}[c]{@{}l@{}}Cosine with \\ Word Vector\end{tabular}}} \\ \hline
\textbf{MultipartiteRank All Sentences}      & 232.4s & 225.1s  & 230.2s \\
\textbf{MultipartiteRank Positive Sentences} & 143.6s & 140.86s & 142.7s \\
\textbf{KEA All Sentences}                   & 97.1s  & 96.28s  & 96.65s \\
\textbf{KEA Positive Sentences}              & 93.5s  & 92s     & 91.72s \\ \hline
\end{tabular}
\end{table}

\begin{figure}[!htbp] 
	\centering
	\subfloat[Word cloud of the keywords extracted by supervised technique KEA.\label{1a}]{%
		\includegraphics[width=0.49\linewidth]{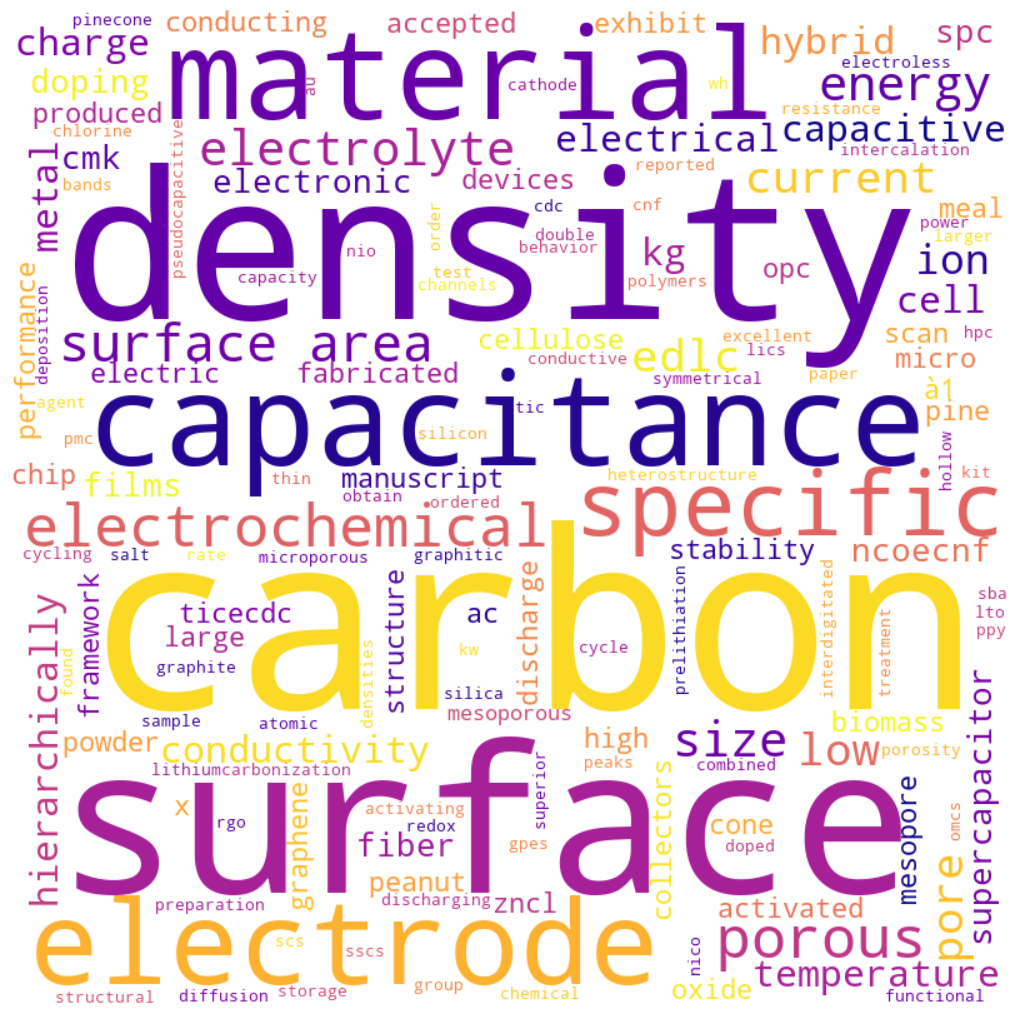}}
	\hfill
	\subfloat[Word cloud of the keywords extracted by unsupervised technique MultipartiteRank.\label{1b}]{%
		\includegraphics[width=0.49\linewidth]{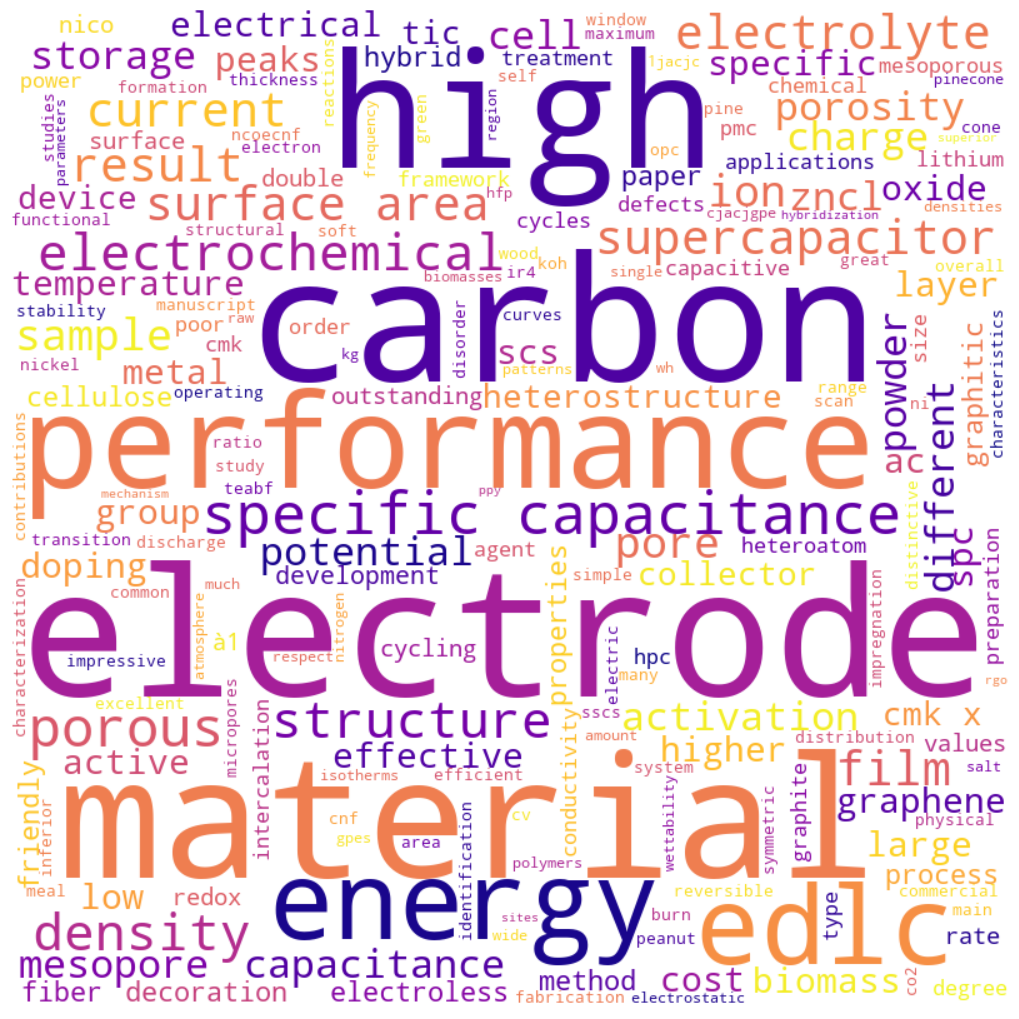}}
	\hfill
	\subfloat[Word cloud of the keywords provided by EDLC domain expert. \label{1c}]{%
		\includegraphics[width=0.58 \linewidth]{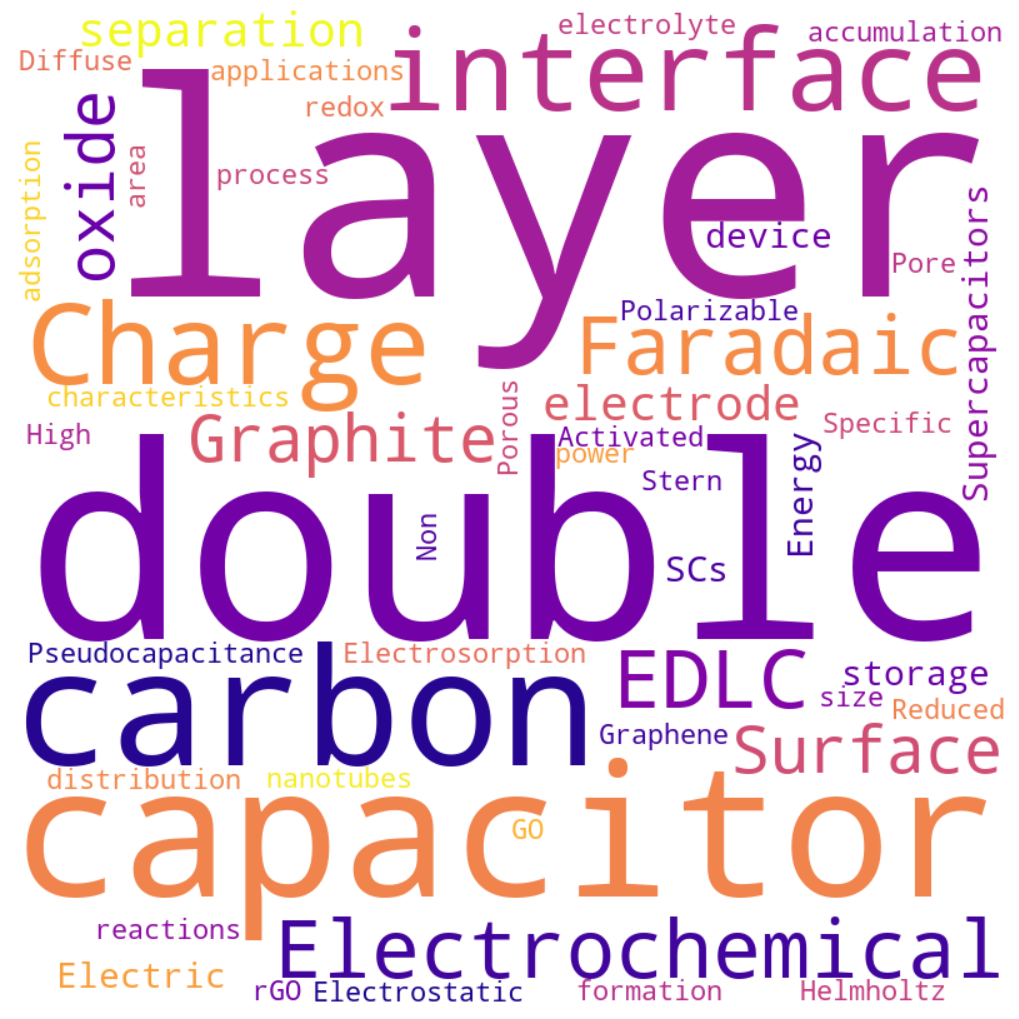}}
	
	\caption{Word cloud representation of the keywords extracted by the top performing keyword extraction techniques achieved with Cosine with word vector similarity index. }
	\label{fig:wordcloud}
\end{figure}

Table \ref{tab:samp_kp} provides the set of keywords extracted by the top-performing keyword extraction techniques employing the Cosine with Word Vector similarity index and expert provided keywords. This table also provides a visual comparison of the similarity between all the keywords. Word cloud representation is also provided in Figure \ref{fig:wordcloud}. Word cloud is utilized to represent the words emphasized according to their frequency, rank, or similarity. This word cloud is generated based on the frequency scores of keywords among all the documents. From the word clouds of top-performing two methods, it is also visible that there are similar keywords of the same scores among all machine-generated and expert provided keywords.  

\begin{table}[!htbp]
	\caption{Sample Keywords extracted by MultipartiteRank, KEA keyword extraction techniques and domain expert curated keywords}
	\label{tab:samp_kp}
	\resizebox{\columnwidth}{!}{\begin{tabular}{|l|l|l|}
			\hline
			\multicolumn{1}{|c|}{\textbf{\begin{tabular}[c]{@{}c@{}}Domain expert \\ curated Keywords\end{tabular}}} & \multicolumn{1}{c|}{\textbf{\begin{tabular}[c]{@{}c@{}}MultipartiteRank \\ extracted \\ Keywords\end{tabular}}} & \multicolumn{1}{c|}{\textbf{\begin{tabular}[c]{@{}c@{}}KEA extracted\\ Keywords\end{tabular}}} \\ \hline
			\begin{tabular}[c]{@{}l@{}}supercapacitors, scs,\\ electrochemical capacitors,\\ energy storage device,\\ electric double\\ layer capacitor,\\ edlc, pseudocapacitance,\\ electrostatic adsorption,\\ electrosorption,\\ faradaic redox reactions,\\ stern layer, helmholtz \\ double layer,\\ double layer   formation,\\ activated carbon ,\\ porous carbon ,carbon \\ nanotubes,\\ graphene, graphite oxide,\\ go, reduced graphite \\ oxide ,rgo, surface charge   \\ accumulation,\\ high power applications,\\ charge separation at \\ electrode interface,\\ charge separation at \\ electrolyte interface,\\ non-faradaic   process,\\ specific surface area,\\ pore size distribution,\\ electrochemical   interface,\\ edlc characteristics,\\ diffuse double layer,\\ polarizable capacitor \\ electrode\end{tabular} & \begin{tabular}[c]{@{}l@{}}layer, power, scs, \\ charge, formation, \\ high, energy, chemical, \\ graphene, surface area, \\ porous carbon, ions, \\ electrolyte, rgo, graphite, \\ energy storage, carbon, \\ electrochemical, surface, \\ pore size distribution, \\ electrode, edlc, \\ supercapacitor, adsorption, \\ supercapacitors, device, \\ capacitance\end{tabular} & \begin{tabular}[c]{@{}l@{}}scs, charge, pore, \\ energy, redox, size, \\ chemical, graphene, \\ ion, surface area, \\ porous carbon, ions, \\ electrolyte, pore size, \\ rgo, graphite, \\ energy storage, carbon, \\ electrochemical, surface, \\ electrode, edlc, \\ specific surface, \\ supercapacitor, porous, \\ specific surface area, \\ oxide, supercapacitors, \\ electric, double, tic, \\ capacitance\end{tabular} \\ \hline
	\end{tabular}}
\end{table}

The study of the experimental results suggests that for extracting keywords and checking the similarity of the extracted keywords from scientific documents, especially for the EDLC related documents, the unsupervised keyword extraction technique MultipartiteRank algorithm can be considered in addition to the expert curated keywords. Although this algorithm requires slightly more computation time than the supervised keyword extraction technique KEA, it gives better results than KEA. If computation time is considered or required over better similarity score, then it is recommended to employ the supervised keyword extraction technique KEA for 1\% of similarity score drop over MultipartiteRank algorithm. When choosing between the positive and the whole article text content, it is recommended to choose the positive text as it has a very small impact on the similarity score but a larger impact on the computation time. Positive texts have no or very little impact on the similarity scores, but require less computation time than all the texts of the scientific articles.


\section{Conclusion}
\label{sec:con}
The aim of this study is to find out which keyword extraction technique provides more similar keywords to the expert provided keywords, which text types have more similarity, which similarity index provides more similarity scores and whether the use of machine generated keywords is feasible with respect to the expert provided keywords. The experiment shows that the unsupervised keyword extraction technique MultipartiteRank provides 92\% similarity with the expert provided keywords in cosine with Word Vector similarity index for positive sentences of the documents from EDLC domain. This study can be further extended with keywords for other domains with a larger dataset in other environments, including author-supplied keywords.

\section*{Data Availability}
Dataset used in this study is available upon request and the request repository is mentioned in Section \ref{subsec:dp}

\section*{Conflicts of Interest}
The authors declare no conflicts of interest.

\section*{Acknowledgment}

Authors would like to express their gratitude to the domain experts for their support and knowledge.

\bibliographystyle{ieeetr}
\bibliography{main}

\begin{thebibliography}{10}
\bibitem{cosine1}
``{Cosine Similarity - Understanding the math and how it works? (with
  python)}.''
\end{thebibliography}

\end{document}